\documentclass[twoside]{article}
\usepackage{amsmath,amscd,amssymb}
\usepackage{qic}

\usepackage{amsthm}
\usepackage{latexsym,enumerate,graphicx,rotating,braket,mathrsfs,url}
\usepackage{color}

\numberwithin{equation}{section}

\theoremstyle{plain}
\newtheorem{thm_}{Theorem}[section]
\newtheorem{lemma_}[thm_]{Lemma}
\newtheorem{prop_}[thm_]{Proposition}
\newtheorem{cor_}[thm_]{Corollary}
\newtheorem{eg_}[thm_]{Example}
\newtheorem{con_}[thm_]{Conjecture}
\newtheorem*{cons_}{Conjecture}

\theoremstyle{definition}
\newtheorem{thmu_}[thm_]{Theorem}
\newtheorem*{thmus_}{Theorem}
\newtheorem{propu_}[thm_]{Proposition}
\newtheorem*{propus_}{Proposition}
\newtheorem{coru_}[thm_]{Corollary}
\newtheorem{lemu_}[thm_]{Lemma}
\newtheorem{egu_}[thm_]{Example}
\newtheorem*{egus_}{Example}
\newtheorem{def_}[thm_]{Definition}
\newtheorem*{defs_}{Definition}
\newtheorem{rk_}[thm_]{Remark}

\newcommand{\thm}[1]{\begin{thm_}#1\end{thm_}}

\newcommand{\lemm}[1]{\begin{lemma_}#1\end{lemma_}}
\newcommand{\lemu}[1]{\begin{lemu_}#1\end{lemu_}}

\newcommand{\prop}[1]{\begin{prop_}#1\end{prop_}}
\newcommand{\propu}[1]{\begin{propu_}#1\end{propu_}}

\newcommand{\defi}[1]{\begin{def_}#1\end{def_}}

\newcommand{\rk}[1]{\begin{rk_}#1\end{rk_}}

\newcommand{\coru}[1]{\begin{coru_}#1\end{coru_}}

\newcommand{\pf}[1]{\begin{proof}#1\end{proof}}

\DeclareMathOperator{\supp}{supp}

\newcommand{\CC}{\mathbb C}

\newcommand{\FF}{\mathbb F}

\newcommand{\tm}{\times}%
\newcommand{\otm}{\otimes}

\newcommand{\lra}{\longrightarrow}

\newcommand{\lmpt}{\longmapsto}

\newcommand{\eq}[1]{\begin{equation}#1\end{equation}}
\newcommand{\eqn}[1]{\begin{equation*}#1\end{equation*}}

\newcommand{\gan}[1]{\begin{gather*}#1\end{gather*}}
\newcommand{\al}[1]{\begin{align}#1\end{align}}
\newcommand{\aln}[1]{\begin{align*}#1\end{align*}}

\newcommand{\enmt}[1]{\begin{enumerate}#1\end{enumerate}}

\newcommand{\cd}{\CC^d\otimes\CC^d}
\newcommand{\ckd}{\CC^d\otimes\CC^{kd}}
\newcommand{\cM}{\mathcal M}
\newcommand{\cU}{\mathcal U}

\begin{document}
\setlength{\textheight}{8.0truein}    

\runninghead{New
 bounds of Mutually unbiased maximally entangled bases in
$\ckd$}{Xiaoya Cheng and Yun Shang }





\fpage{1}

\centerline{\bf
New bounds of mutually unbiased maximally entangled bases in
$\ckd$}

\vspace*{0.37truein}
\centerline{\footnotesize
XIAOYA CHENG}
\vspace*{0.015truein}
\centerline{\footnotesize\it Institute of Mathematics,
Academy of Mathematics and System Science,
CAS
}
\centerline{\footnotesize\it School of Mathematical Sciences, University of Chinese Academy of Sciences}
\baselineskip=10pt
\centerline{\footnotesize\it Beijing 100190, P.R. China
}
\vspace*{10pt}
\centerline{\footnotesize
YUN SHANG
\footnote{Corresponding author. Emai: shangyun@amss.ac.cn}}
\vspace*{0.015truein}
\centerline{\footnotesize\it Institute of Mathematics,
Academy of Mathematics and System Science,
CAS\\}
\centerline{\footnotesize\it Key Laboratory of Management, Decision and Information Systems, CAS\\}
\centerline{\footnotesize\it National Center for Mathematics and Interdisciplinary Sciences, CAS
}
\baselineskip=10pt
\centerline{\footnotesize\it Beijing 100190, P.R. China
}


\vspace*{0.21truein}

\abstracts{
\noindent\textbf{Abstract}
Mutually unbiased bases which is also maximally entangled bases is called mutually unbiased maximally entangled bases (MUMEBs). We study the construction of MUMEBs in bipartite system. In detail, we construct $2(p^a-1)$ MUMEBs in $\cd$ by properties of Gauss sums for arbitrary odd $d$. It improves the known lower bound $p^a-1$ for odd $d$. Certainly, it also generalizes the lower bound $2(p^a-1)$ for $d$ being a single prime power.
Furthermore, we construct MUMEBs in $\ckd$ for general $k\geq 2$ and odd $d$. We get the similar lower bounds as $k,b$ are both single prime powers.
Particularly, when $k$ is a square number, by using mutually orthogonal Latin squares, we can construct more MUMEBs in $\ckd$, and obtain greater lower bounds than reducing the problem into prime power dimension in some cases.
}{}{}

\vspace*{10pt}

\keywords{mutually unbiased bases, maximally entangled states, Pauli matrices, mutually orthogonal Latin squares}
\vspace*{3pt}

\vspace*{1pt}\textlineskip    

\section{Introduction}\label{sec_intro}

Complementarity is the fundamental concept of quantum theory, which means that there exist observables
that cannot be measured simultaneously. This phenomenon is most strongly
manifested when observables are related to mutually unbiased bases(MUBs). Two
bases are said to be unbiased if all (normalized) eigenvectors
of one observable have the same overlap with
all eigenvectors of the other observable.
According, if a system
is in an eigenstate of a particular basis, then the
measurement result in a corresponding MUBs is completely uncertain. MUBs find many applications in quantum information task such as quantum error correction codes \cite{pawlowski2010entanglement}, quantum state tomography
\cite{ivonovic1981geometrical, wootters1989optimal}, quantum key distribution \cite{cerf2002security}, cryptographic protocols \cite{cerf2002security,brierley2009quantum}, mean king problem \cite{aharonov2001mean}, quantum teleportation and superdense coding \cite{durt2004if, sangare2009continuum, revzen2010maximally}.

How many MUBs exist for any dimension Hilbert space is still an open problem. A recent review can refer to \cite{durt2010mutually}.
In general, for $d\geq 2$, it is proved that $p_1^{a_1} +1\leq N(d)\leq d+1$ for $d= p_1^{a_1}\dots p_s^{a_s}$ with $p_1^{a_1}\le\dots\le p_s^{a_s}$, where $N(d)$ denotes the maximum number of MUBs in the $d$-dimensional Hilbert space $\mathbb{C}^d$. And $N(d)=p^{a}+1=d+1$ when $d$ is a single prime power \cite{wootters1989optimal}. That is, it is possible to find $d + 1$ MUBs, which is called a complete set of MUBs. There are many different methods to construct MUBs. By Weil sums over finite fields and exponential sums over Galois rings, Klappenecker et al. \cite{klappenecker2004constructions} studied MUBs for odd prime power $d=p^a, p\geq 3$ and even prime power $d=2^m$ respectively.
Wocjan et al. \cite{wocjan2004new} showed that
for $d=s^2$ the number of $N(d)$ is greater than $s^{\frac{1}{14.8}}$ for all $s$
but finitely many exceptions by orthogonal Latin square. Obviously this bound is better than the previous one in many non-prime-power cases. But if $d$ is a composite number, the value of $N(d)$ is still unknown.

When the vector space is a bipartite system $\CC^d\otimes\CC^{d'}$ of composite dimension $dd'$,
there are different kinds of bases in according to the entanglement of the basis vectors, such as unextendible product basis\cite{Benett 1999}, unextendible maximally entangled basis \cite{Chen 2013} and maximally entangled basis\cite{tao2015mutually}. A basis $\mathcal B$ of $\CC^d\otimes\CC^{d'}$ is called a maximally entangled basis (MEB)
 if it consists of $dd'$ maximally entangled states. Maximally entangled states is a very important concept in quantum information science. It plays a vital role in quantum computing and quantum communication tasks, such as measurement based quantum computing, quantum key distribution, quantum teleportation etc. Certainly, entanglement is always present in a complete
set of MUBs \cite{Wie¡äsniak 2011entanglement}. So discussing mutually unbised bases which are also maximally entangled basis become a new interesting topic recently. Let $M(d,d')$ be the maximal cardinality of any set of MUMEBs in $\CC^d\otimes\CC^{d'}$. Tao et al. proved that $M(2,4)\ge 5$ and $M(2,6)\ge 3$ in \cite{tao2015mutually}. Liu et al. constructed $p_{1}^{a_{1}}-1$ MUMEBs,
i.e., $M(d,d)\ge p^{a_1}_1-1$ for arbitrary $d\ge 2$.
 Here $d=p_1^{a_1}\dots p_s^{a_s}$ with $p_1^{a_1}\le \dots\le p_s^{a_s}$, $p_1,p_2,\cdots,p_s$
  are distinct primes \cite{liu2017mutually}. In \cite{xu2017construction}, Xu showed that if $d$ is a single prime power, that is $d=p^{a}$, then it is possible to find $2(p^{a}-1)$ MUMEBs. Furthermore, Xu constructed MUMEBs in $\ckd$ when $k$ is also a single prime power and obtained $M(d,kd)\ge \min\{k,M(d,d)\}$.

However, the problem to find the lower bound on $M(d,kd)$ for more general $d$ and $k$
 remains unknown.

In this paper, we will focus on constructing new lower bounds of $M(d,kd)$ for more general $k$ and odd $d$ in $\ckd$.
We constructed $2(p_1^{a_1}-1)$ MUMEBs in $\cd$ by properties of Gauss sums (reference to Proposition \ref{prop_gauss}
and Proposition \ref{lem_gauss_sum}
) for arbitrary odd $d$, when $k=1$. It improves the known
lower bound $p_1^{a_1}-1$ for odd $d$ \cite{xu2017construction}. Certainly, it also generalizes the lower bound $2(p_1^{a_1}-1)$ from $d$ being
a single prime power into a generic odd $d$ \cite{liu2017mutually}. Furthermore, by eliminating the restriction on $d$ and $k$
to be prime powers in \cite{xu2017construction}, we constructed MUMEBs in $\ckd$ for
general $k\geq 2$ and odd $d$. We got the similar lower bounds as $k,b$ are both single prime powers in \cite{xu2017construction}, that is $M(d,kd)\ge \min\{(p'_1)^{a'_1}+1,M(d,d)\}$.

Especially, since Latin square is also a useful tool in characterizing MUBs problem (c.f. \cite{hall2011mutually}) in single system,
we first consider whether mutually orthogonal Latin square (MOLS)(c.f. \cite{wocjan2004new})
is helpful to improve the value of $M(d,kd)$ in bipartite system.
By using results on MOLSs in \cite{wocjan2004new},
 we obtained some new results
 on the lower bound for $M(d,kd)$ (reference to Theorem \ref{thm_sq}
). We found that $M(d,kd)\ge \min\{N_{\text{MOLS}}(\sqrt{k})+2,M(d,d)\}$ for $k$ being a square number, where $N_{MOLS}(x)$ denotes the maximum cardinality of any set of MOLS of order $x$ (see \cite{wocjan2004new}). We also discuss the relation between the above two kinds of lower bounds. If $k=p_1^{2e_1}$ is a square of a prime power,
  the first bound of $M(d,kd)$ is better than the second one.
 But if $l\ge 35$, the second one is better. Anyway, assuming $k$ is any square number, we have
$M(d,kd)\ge \min\{ \max\{N_{\text{MOLS}}(\sqrt{k})+2,(p'_1)^{a'_1}+1\},M(d,d)\}$.

In addition, all the bounds that we obtain in this paper still hold for general $k$ and $d=2^m$.

The paper is organized as follows.
In Section \ref{sec_basic}, we give some basic definitions and review a basic criterion of MUMEBs in $\ckd (d\geq 2)$.
Section \ref{sec_cd} is devoted to the construction of MUMEBs in $\cd$  ($k=1$) for odd $d$ and the lower
 bound for $M(d,d)$ by using unitary matrices and properties of Guss sums.
In Section \ref{sec_ckd}, we consider the same problem in $\ckd$, where $k\ge 2$ and $d$ is
 odd.
 Note that in \cite{xu2017construction}, $k$ and $d$ are only restricted to prime powers.
In Section \ref{sec_square}, when $k$ is a square number, by using mutually orthogonal Latin squares, we
construct more MUMEBs in $\ckd$, and obtain greater lower bounds than reducing the problem
into prime power dimension in some cases.
In Section \ref{sec_con}, we give conclusions and raise some future problems.

\section{A basic criterion for MUMEBs in $\ckd$}\label{sec_basic}
We introduce the general construction and criterion for MUMEBs in $\ckd$ \cite{xu2017construction}.
Let $R$ be a commutative ring with $1$ and $R^*$ be the group of invertible elements in $R$.
Let $\cM_n(\CC)$ denote the ring of $n\tm n$ matrices over complex number field $\CC$,
and $\cU_n(\CC)$ be the group of unitary matrices in $\cM_n(\CC)$, where
$I_n$ is the unit matrix in $\cM_n(\CC)$ and $U^{\dagger}$ is the transpose conjugation of $U\in \cM_n(\CC)$.
\defi{
A pure state $\ket\Psi$ is said to be a \emph{maximally entangled state} in $\mathbb{C}^d\otimes \mathbb{C}^{d'}$
($d\le d'$) if and only if for an arbitrary given orthonormal
complete basis $\{\ket{\phi_i}\}_{i=1}^d$ of subsystem A, there exists an orthonormal basis $\{\ket{\psi_j}\}_{j=1}^{d'}$ of subsystem B such that
$$\ket\Psi=\frac{1}{\sqrt d}\sum_{i=0}^{d-1}\ket{\phi_i}\otimes\ket{\psi_i}.$$
}
\defi{ Two orthogonal bases $\mathcal{B}_1=\{\ket{\phi_i}\}_{i=1}^d$ and $\mathcal{B}_2=\{\ket{\psi_j}\}_{j=1}^d$ of $\mathbb{C}^d$ are called \emph{mutually unbiased} if
$$|\langle \phi_i|\psi_j\rangle|=\frac{1}{\sqrt{d}},\ \ \ \ (1\leq i, j\leq d).$$
A set of orthonormal bases $\mathcal{B}_1, \mathcal{B}_2, \ldots, \mathcal{B}_m$ in $\mathbb{C}^d$ is said to be a set of MUBs if every pair of $\mathcal{B}_i$ and $\mathcal{B}_j$ $(1\leq i\neq j\leq d)$ in the set is mutually unbiased.}
\defi{
An \emph{additive character} of $R$ is a homomorphism from the additive group $R$
to the multiplicative group  $\CC^*$.
A \emph{generic character} of $R$ is an additive character
$\lambda: R\lra \CC^*$ such that $\sum_{r\in R}\lambda(ar)=0$ for all
 $a\in R\setminus\set{0}$.
}

Assume that there exists a generic character $\lambda$ of $R$.
We also fix an orthonormal basis
 $$\set{\ket{e_r}|r\in R}$$
 of  $\CC^d$  indexed by $R$ and an orthonormal basis
 $$\set{\ket{e'_{r,j}}|r\in R, j=1,2,\dots,k}$$
 of $\CC^{kd}$  indexed by $R\tm \set{1,2,\dots,k}$.

Given  $U$ in $\cU_{kd}(\CC)$,  we consider the
 following $k$ maximally entangled states in $\ckd$
 $$\ket{\psi_U^j}=\frac{1}{\sqrt d}\sum_{r\in R} \ket{e_r}\otm U\ket{e'_{r,j}},
    \quad j=1,2,\dots,k.$$
Define \emph{Pauli operators}
 $$H_{\xi,\eta}=\sum_{r\in R}\lambda(r\xi)\ket{e_{r+\eta}}\bra{e_r}, \quad \xi,\eta\in R$$
  Applying $H_{\xi,\eta}\otm I_{kd}$ on $\ket{\psi_U^j}$,
 then we obtain the following $kd^2$ maximally entangled states
\gan{
(H_{\xi,\eta}\otm I_{kd})\ket{\psi_U^j} = \frac{1}{\sqrt{d}}\sum_{r\in R}
    \lambda(r\xi)\ket{e_{r+\eta}}\otm U\ket{e'_{r,j}},\\
\xi,\eta\in R, j=1,2,\dots,k.
}
Set
 $$\Psi_U=\set{(H_{\xi,\eta}\otm I_{kd})\ket{\psi_U^j}| \xi,\eta\in R, j=1,\dots,k}.$$
 Then we have the following basic criterion for these maximally entangled bases
 being mutually unbiased.
\propu{[See \cite{xu2017construction}] \label{prop_cr}
Let notations be as before. We have the following results.
\enmt{[(1)]
\item For any $U$ in $\cU_{kd}(\CC)$, $\Psi_U$ is an orthonormal maximally entangled basis (MEB)
 in $\ckd$.
\item For $U$ and $V$ in $\cU_{kd}(\CC)$, $\Psi_U$ and $\Psi_V$
 in $\ckd$ are mutually unbiased if and only if
 \eq{\label{eq_crk}
  \left| \sum_{r\in R} \lambda(r\xi)w_{(r,j),(r+\eta,l)} \right| =\frac{1}{\sqrt k},
       \quad  \text{ for all } \xi,\eta\in R \text{ and } j,l=1,\dots,k,
  }
  where $U^\dagger V=(w_{(r,j),(s,l)})$, $(r,j),(s,l)\in R\tm \set{1,\dots,k}$.
 In particular, if $k=1$, then $\Psi_U$ and $\Psi_V$  in $\cd$ are mutually unbiased if and only if
 \eq{\label{eq_cr1}
  \left| \sum_{r\in R} \lambda(r\xi)w_{r,r+\eta} \right| =1,
       \quad  \text{ for all } \xi,\eta\in R.
  }
}}

\section{Construction of MUMEBs in $\cd$}\label{sec_cd}
In this section, we restrict ourself to the case  $\cd$, i.e., $k=1$.
By Proposition \ref{prop_cr}, we see that $\Psi_U$ and $\Psi_V$
 in $\cd$ are mutually unbiased, provided that  $U$ and $V$ in $\cM_{d}(\CC)$
 satisfy \eqref{eq_cr1}.
Thus we need only to construct a set of matrices in $\cU_{d}(\CC)$,
 such that they satisfy \eqref{eq_cr1} in pair.
Suppose  we have the decomposition $d=p_1^{a_1}\dots p_s^{a_s}$
 with $p_1^{a_1}\le \dots\le p_s^{a_s}$,
 where $p_t, t=1,2,\dots,s$ are distinct primes.
As mentioned in the introduction, Liu et al. \cite{liu2017mutually}
 constructed a set of permutation matrices $\set{U}$, having size $p_1^{a_1}-1$,
 and thus showed that $M(d,d)\ge p_1^{a_1}-1$.
Note here $p_1^{a_1}$ is actually  the minimal prime power dividing $d$.

After that, Xu \cite{xu2017construction} restricted $d=p^a$ to be a single prime power
 and constructed another set of unitary matrices $\set{V}$
 from  $\set{U}$, reaching a better lower bound $M(d,d)=M(p^a,p^a)\ge 2(p^a-1)$.

What's the result for general $d$? By introducing quadratic Gauss sums, we prove that $M(d,d)\ge 2(p_1^{a_1}-1)$ for any
 odd $d=p_1^{a_1}\dots p_s^{a_s}$,
 with $p_1^{a_1}\le \dots\le p_s^{a_s}$.

 Our construction process is similar to Xu's \cite{liu2017mutually}.
Let us first recall the construction in \cite{liu2017mutually}.

For each $a\in R^*$, define $U(a)$ by
\eqn{
U(a)=(u(a))_{r,s\in R},\ \ u(a)_{r,s}=\delta_{ar,s}\text{ for all } r,s\in R.
}
For each $r\in R$,
$U(a)\ket{e_r}=\sum_{l\in R}u(a)_{lr}\ket{e_l}=\ket{e_{a^{-1}r}}$,
which shows that $U(a)$ is a permutation matrix.
Actually, $U$ induces a monomorphism
$$U: R^*\lra \cU_d(\CC).$$
In particular,  for all $a,b\in R^*$ we have
\enmt{[(1)]
\item $U(a)=I_d$ if and only if $a=1$,
\item $U(a)U(b)=U(ab)$,
\item $U(a)^\dagger=U(a^{-1})=U(a)^{-1}$.
}

Next we describe the further construction in \cite{xu2017construction}.
Define  $W\in\cU_d(\CC)$ as follows:
\eqn{
W_{r,s}=\frac{1}{\sqrt d}\lambda(rs) \text{ for all } r,s\in R,
}
Then  for each   $a\in R$, set $V(a)=U(a)W\in\cU_d(\CC)$.

\lemm{\cite{xu2017construction}\label{lem_uv}
We have the following statements:
\enmt{[(1)]
\item \label{it_u} For any $a,b\in R^*$, $\Psi_{U(a)}$ and $\Psi_{U(b)}$ are
 mutually unbiased, provided that $a-b\in R^*$.
\item \label{it_v} For any $a,b\in R^*$, $\Psi_{V(a)}$ and $\Psi_{V(b)}$ are
 mutually unbiased, provided that $a-b\in R^*$.
\item \label{it_1v} For any $a\in R^*$, $\Psi_{I_d}$ and $\Psi_{V(a)}$ are
 mutually unbiased, provided that
\eq{\label{eq_squaresum}
\text{$2\in R^*$ and }
\lambda\text{ satisfies }
\left|\sum_{r\in R} \lambda(cr^2)\right| = \sqrt{d}
\text{ for any $c\in R^*$}.
}
\item \label{it_uv} For any $a,b\in R^*$, $\Psi_{U(a)}$ and $\Psi_{V(b)}$ are
 mutually unbiased, provided $\lambda$ satisfies \eqref{eq_squaresum}.
}}
\pf{
The basic idea is to use  Proposition \ref{prop_cr}.
The proof of \eqref{it_u} is originally given by \cite[Lemma~3.1]{liu2017mutually}.
For \eqref{it_v}, note that  $b^{-1}a-1 = b^{-1}(a-b)\in R^*$. The following argument is
 the same as in \cite[Corollary~3.2]{xu2017construction}.
As for \eqref{it_1v}, we only have the following equation by the proof of \cite[Lemma~3.3]{xu2017construction},
$$ \sqrt{d} \left|\sum_{r\in R} \lambda(rx)(V(a)_{r,r+y})\right|
    = \left|\sum_{r\in R} \lambda(ar^2+(ay+x)r)\right|.$$
Now, we give a supplementary proof,
i.e., $\left|\sum_{r\in R} \lambda(ar^2+(ay+x)r)\right| = \sqrt d$.
Noting the assumption \eqref{eq_squaresum}, we have
\aln{
\left|\sum_{r\in R} \lambda(ar^2+(ay+x)r)\right|
    &= \left|\sum_{r\in R} \lambda(a(r+(ay+x)/2a)^2-(ay+x)^2/4a)\right|\\
    &= \left|\sum_{r\in R} \lambda(a(r+(ay+x)/2a)^2)\lambda(-(ay+x)^2/4a)\right|\\
    &= \left|\sum_{r\in R} \lambda(a(r+(ay+x)/2a)^2)\right|\left|\lambda(-(ay+x)^2/4a)\right|\\
    &= \left|\sum_{r\in R} \lambda(ar^2)\right| = \sqrt d.
}
Then Proposition \ref{prop_cr} yields \eqref{it_1v}.
At last, \eqref{it_uv} is a corollary of \eqref{it_1v} and we refer the
 reader to \cite[Corollary 4.6]{xu2017construction}.
The proof is complete.
}

To show the existence of MUMEBs in $\cd$ for odd $d$, we need to specify the individual
 $R$ and $\lambda$. We use the same construction for $R$ and $\lambda$ in Liu et al. \cite{liu2017mutually}.
Specifically, let $d=p_1^{a_1}\dots p_s^{a_s}$ with $p_1^{a_1}\le \dots\le p_s^{a_s}$,
 $q_t = p_t^{a_t}$ and $R=\FF_{q_1}\oplus \dots \oplus \FF_{q_s}$. Note that $d$ is odd,
 so the character of $R$ is not $2$.
Then clearly $\left|R\right|=q_1q_2\dots q_s = d$.
Define
\al{\label{eq_lambda}
\lambda: R        &\lra     \CC^*\\
(x_1,\dots,x_s)   &\lmpt    \prod_{t=1}^s\zeta_{p_t}^{T_{\FF_{q_t}/\FF_{p_t}}(x_t)},\nonumber
}
where $T_{\FF_{q_t}/\FF_{p_t}}$ is the trace map from $\FF_{q_t}$ to $\FF_{p_t}$,
i.e. $T_{\FF_{q_t}/\FF_{p_t}}(x_t)=x_t+x_t^{p_t}+x_t^{{p_t}^2}+\dots +x_t^{p_t^{a_t-1}}$ and $\zeta_{p_t}
=e^{\frac{2\pi i}{p_t}}$.
From the proof of \cite[Theorem 3.3]{liu2017mutually}, we know that
 $\lambda$ is a generic character of $R$.
By restricting $d=p^a$ to be a prime power, Xu \cite{xu2017construction}
 proved that the critical assumption \eqref{eq_squaresum} holds.

However, the following properties of Gauss sum \cite{berndt1998gauss} shows that \eqref{eq_squaresum} holds for any odd $d$.
\propu{\label{prop_gauss}(See \cite[pp. 10--11]{berndt1998gauss}.)
Let    $q=p^a$, $c\in \FF^\times_q$ and $\chi$ a
 multiplicative character of $\FF_q$.
Define
\aln{
G_a(c,\chi)&=\sum_{r\in \FF_q} \zeta_p^{T_{\FF_q/\FF_p}(cr)} \chi(r),\\
g_a(c,k)   &=\sum_{r\in\FF_q}  \zeta_p^{T_{\FF_q/\FF_p}(cr^k)}.
}
Then we have
\enmt{[(1)]
\item $\left|G_a(c,\chi)\right|=\sqrt{q}$,  if $\chi$ is nontrivial,
\item $g_a(c,k) = \sum_{j=1}^{k-1} G_a(c,\chi^j)$ where $\chi$ is a character of order $k$.
}
}
\prop{\label{lem_gauss_sum}
Let $d$ be an odd number.
For the above $R$ and $\lambda$,
we have for any $c\in R^*$, $\left|\sum_{r\in R} \lambda(cr^2)\right| = \sqrt{d}$.
That is, the assumption \eqref{eq_squaresum} holds.
}
\pf{
Suppose $q$ is an odd prime power.
Let $\chi$ be a character of order $2$.
Then by Proposition \ref{prop_gauss},
 we have for any $c\in \FF_q^*$,
$$ \left|\sum_{r\in \FF_q} \zeta_p^{T_{\FF_q/\FF_p}(cr^2)}\right| = \left|g_a(c,2)\right|
     = \left|G_a(c,\chi)\right| = \sqrt{q}. $$

Now let $R=\FF_{q_1}\oplus\dots\oplus \FF_{q_s}$ and
 $c=(c_1,\dots,c_s)\in  R^* = \FF_{q_1}^*\oplus \dots \oplus \FF_{q_s}^*$. Then we have
\aln{
\left|\sum_{r\in R} \lambda(cr^2)\right|
    &= \left|\sum_{(r_1,\dots,r_s)\in \FF_{q_1}\oplus \dots \oplus \FF_{q_s}}
        \lambda((c_1,\dots,c_s)(r_1,\dots,r_s)^2)\right| \\
    &= \left|\sum_{(r_1,\dots,r_s)\in \FF_{q_1}\oplus \dots \oplus \FF_{q_s}}
        \lambda((c_1r_1^2,\dots,c_sr_s^2))\right| \\
    &= \left|\sum_{(r_1,\dots,r_s)\in \FF_{q_1}\oplus \dots \oplus \FF_{q_s}}
        \prod_{t=1}^s \zeta_{p_t}^{T_{\FF_{q_t}/\FF_{p_t}}(c_tr_t^2)}\right|\\
    &= \prod_{r_1\in \FF_{q_1},\dots,r_s\in \FF_{q_s}} \left|
        \sum_{r_t\in \FF_{q_t}} \zeta_{p_t}^{T_{\FF_{q_t}/\FF_{p_t}}(c_tr_t^2)}\right|\\
    &= \prod_{r_1\in \FF_{q_1},\dots,r_s\in \FF_{q_s}} \sqrt{q_t} = \sqrt{d}.
}
}

\thm{\label{thm_cd}
Let $d$ be an odd number.
Write $d=p_1^{a_1}\dots p_s^{a_s}$ with $p_1^{a_1}\le \dots\le p_s^{a_s}$.
Then $M(d,d)\ge 2(p_1^{a_1}-1)$. That is, there exists
 a set of  MUMEBs in $\cd$ of size  $2(p_1^{a_1}-1)$.
}
\pf{
If $d=p_1^{a_1}\dots p_s^{a_s}$ is odd,
let  $R$ and $\lambda$ be in line with  the previous constructions.
Since $q_t-1\geq q_1-1$ for all $t>1$, where $q_t = p_t^{a_t}$, we can fix an injection
$\iota_t:\FF_{q_1}^*\lra \FF_{q_t}^*$ for each $t>1$. Define
\eqn{
S=\set{(u, \iota_2(u), \dots, \iota_s(u))\in R^* | u\in \FF_{q_1}^*}.
}
Clearly, $S$ is a subset of $R^*$ such that $a-b\in R^*$ for all $a\neq b\in S$.
It follows from Lemma \ref{lem_uv} \eqref{it_u} that
 $\set{\Psi_{U(a)}|a\in S}$ is a set of MUMEBs
    and    from Lemma \ref{lem_uv} \eqref{it_v} that
 $\set{\Psi_{V(a)}|a\in S}$ is also a set of MUMEBs.
Moreover, Proposition \ref{lem_gauss_sum} tells us that
 the assumption \eqref{eq_squaresum} holds for the choice of $R$ and $\lambda$.
It follows from  Lemma \ref{lem_uv} \eqref{it_uv} that
  $\Psi_{U(a)}$ and $\Psi_{V(a)}$ are mutually unbiased for any $a\in R^*$.

In summary, $\set{\Psi_{U(a)}|a\in S} \cup \set{\Psi_{V(a)}|a\in S}$ is
 a set of MUMEBs.
In particular, the  size of this set is $2\mid S\mid=2(p_1^{a_1}-1)$.

Thus the proof is compete.
}

\section{Construction of MUMEBs in $\ckd$ for general $k\ge 2$}\label{sec_ckd}
Let $d$ be odd
and $k\ge 2$,
then we have the decomposition $k=(p'_1)^{a'_1}\dots (p'_l)^{a'_l}$
with $(p'_1)^{a'_1}\le \dots\le (p'_l)^{a'_l}$,
 where each $p'_t, t=1,2\dots,l$ is distinct prime.
In line with \cite[Section 5]{xu2017construction},
we construct MUMEBs in $\ckd$. 
But unlike \cite{xu2017construction}, we do not need to restrict $d$ and $k$ to be prime powers.
Therefore we obtain the MUMEBs in $\ckd$ for general $k$ and odd $d$.

For each $t=1,\dots,l$,
 if $p'_t$ is odd, let $q'_t=(p'_t)^{a'_t}$ and for
 $j_t\in \FF_{q'_t}$, define
 $$B_{j_t}^{(t)}=\left(\frac{1}{\sqrt{q'_t}}\zeta_{p'_t}
    ^{T_{\FF_{q'_t}/\FF_{p'_t}}(j_t m^2+mn)}\right)_{(m,n)\in\FF_{q'_t}^2};$$
 if $p'_t=2$, define
 $$B_{j_t}^{(t)}=\left(\frac{1}{\sqrt{2^{a'_t}}}\zeta_4
    ^{(j_t+2n)m}\right)_{(m,n)\in\mathcal T_{a'_t}^2}\quad \forall j_t\in\mathcal T_{a'_t},$$
where $\mathcal T_{a'_t}$ is a set of $2^{a'_t}$ element in the Galois ring $GR(4,a'_t)$
 (see \cite{xu2017construction} for detailed definitions).
By the properties of Gauss sums and Galois rings, and a similar
 argument as in the proof of Proposition \ref{lem_gauss_sum},
 one can check that $B_{j_t}^{(t)}\in\mathcal U_{q'_t}(\CC)$ and the absolute value of
 each entry in $B_{j_t}^{(t)^{\dagger}}B_{i_t}^{(t)}$ equals to $1/\sqrt{ q'_t}$ for
 any two distinct $j_t, i_t\in \FF_{q'_t}$.

 Fix an injection $\nu_t:\FF_{q_1'}\lra\FF_{q_t'}$ for each $t>1$ and
define
\eq{\label{eq_b}
B_j = B_{j}^{(1)}\otm B_{\nu_2(j)}^{(2)}\otm\dots\otm B_{\nu_l(j)}^{(l)}\in\cU_k(\CC),
  \quad j=1,\dots,q_1'
}
We also write $B_0=I$.
By the property of matrix tensor product,
 then $(A\otm B)(C\otm D) = AC\otm BD$,
 one has  $B_j\in\cU_k(\CC)$ and  the absolute value of
 each entry in $B_j^{\dagger}B_i$   equals to $1/\sqrt{k}$ for
 any two distinct $j, i\in \set{0,1,\dots,q'_1}$.

\thm{\label{thm_odd}
Let $d$ be an odd number.
Write $d=p_1^{a_1}\dots p_s^{a_s}$ with $p_1^{a_1}\le \dots\le p_s^{a_s}$.
Suppose $k\ge 2$ and write $k=(p'_1)^{a'_1}\dots (p'_l)^{a'_l}$
with $(p'_1)^{a'_1}\le \dots\le (p'_l)^{a'_l}$.
Then $$M(d,kd)\ge \min\{(p'_1)^{a'_1}+1,M(d,d)\}\ge \min\{(p'_1)^{a'_1}+1, 2(p_1^{a_1}-1)\}.$$
}
\pf{
Let $d$ be an odd number.
Without loss of generality, suppose that $(p'_1)^{a'_1}+1\le M(d,d)$,
 since otherwise we can prove the result similarly.
Let $n=(p'_1)^{a'_1} = q'_1$ and $\{ B_j|j=0,1,\dots,n\}$ be defined as before.
Since $n+1\le M(d,d)$, by Theorem \ref{thm_cd}
there exist $U_0,U_1,\dots,U_n$  distinct matrices in $\cU_d(\CC)$
 such that $\Psi_{U_0},\Psi_{U_1},\dots,\Psi_{U_n}$ are MUMEBs.
For any $0\le t\le n$, $C_t=B_t\otimes U_t$
 is a unitary matrix. The following prove that these MEBs $\{\Psi_{C_i}\}_{i=1}^n$ are mutually unbiased i.e.,
 for any $0\le t<t'\le n$, the matrix $C_t^{\dagger}C_{t'}$ satisfies \eqref{eq_crk}.
Let $\xi,\eta\in\FF_d$ , $0\le t<t'\le n$
and $B_t^{\dagger}B_{t'}=(b_{i,j})$, $U_t^{\dagger}U_{t'}=(u_{i,j})$. Then
\aln{
\left| \sum_{r\in R} \lambda(r\xi)(C_t^{\dagger}C_{t'})_{(r,i),(r+\eta,j)} \right|
     &=\left| \sum_{r\in R} \lambda(r\xi)b_{i,j}u_{r,r+\eta} \right|\\
     &=\left| b_{i,j}\sum_{r\in R} \lambda(r\xi)u_{r,r+\eta} \right|\\
     &=\left| b_{i,j}\right|\left|\sum_{r\in R} \lambda(r\xi)u_{r,r+\eta} \right|\\
     &=\frac{1}{\sqrt k},
}
The last equality follows from
 $b_{i,j}=1/\sqrt{ k}$  and \eqref{eq_crk}.

The proof is complete.
}

\section{Construction of MUMEBs in $\ckd$ with $k$ being a square number}\label{sec_square}
In the previous sections, we obtain a  bound for $M(d,kd)$ for general $k$.
Now we consider it for some special $k$.
It turns out that if $k$ is a square number, the bound for $M(d,kd)$ can be improved.

Since the problem to determine $N(d)$ is similar to the combinatorial
 problem to determine the maximal size $N_{\text{MOLS}}(d)$ of
 all sets of mutually orthogonal Latin squares  (MOLSs)
 of size $d\times d$ \cite{hall2011mutually}, many people studied the problem from the point of view of Latin square,
 such as Klappenecker \cite{klappenecker2004constructions}, Musto \cite{musto2016constructing},
 Rao \cite{rao2010mutually} and so on.
Wocjan et al. \cite{wocjan2004new} gave the construction of MUBs
 in square dimensional case by using orthogonal Latin squares.
 They gave more mutually orthogonal bases in many non-prime-power
 dimensions by using some kind of net as a bridge.

In this section, we generalize this idea to bipartite system $\ckd$
 and obtain more MUMEBs in $\ckd$ in some cases.

Suppose that $k$ is a square number, like $k=x^2$.
We begin with some necessary definitions (c.f. \cite{wocjan2004new}).

\defi{ Let $m:=(m[1],\dots,m[k])^T$ be a column vector of size $k$.
If its entries take only the values $0$ and $1$, i.e., $m\in \{0,1\}^k$,
the vector $m$ is called an \emph{incidence vector}.
}
Notes that the \emph{Hamming weight} of $m$ is the number of $1$'s.
Denote the support of incidence vector $m$ as $\supp(m)$.
Then $\supp(m)=\{j_1,\dots,j_x\}$, where $j_i$'s are all indices such that
 the corresponding entries $m[j_1],\dots,m[j_x]$ of $m$ are all $1$
 (the Hamming weight of $m$ is $x$).
\defi{[$(n,x)$-net] Let $\{m_{11},\dots,m_{1x},m_{21},\dots,m_{2x},\dots,m_{n1},\dots,m_{nx}\}$ be a collection
 of incidence vectors of size $k=x^2$ that are partitioned into $n$ blocks and each block contains $x$
 incidence vectors. Let $m_{bi}$ denote the incidence vectors, where $b\in \{1,\dots,n\}$ identifies the block
 and $i\in\{1,\dots,x\}$ the vector within a block. We say that the incidence vectors form a $(n,x)$\emph{-net} when the following conditions holds:
\enmt{[(1)]
\item The supports of all vectors are disjoint in the same block, i.e.,$$m^T_{bi}m_{bj}=0$$
for all $b\in\{1,\dots,n\}$, $1\le i\neq j\le x$.
\item The intersection of any incidence vectors from two different blocks contains exactly one element ,i.e.,
$$m^T_{bi}m_{b'j}=1$$ for all $1\le b\neq b'\le n$, $1\le i, j\le x$.
}
}
\defi{
Let $m\in \{0,1\}^k$ be an incidence vector of Hamming weight $x$ and $h\in \CC^x$ be an arbitrary column vector.
 Then $h\uparrow m$ denotes \emph{the embedding of $h$ into $\CC^k$ controlled by $m$}, to be the following vector in $\CC^k$
 $$h\uparrow m:=\sum_{i=1}^xh[i]\ket{j_i},$$
 where $h[i]$ is the $i$th entry of the vector $h$,
 $\{j_1,\dots,j_x\}$ is the support of $m$ with the ordering $j_1<j_2<\dots<j_x$ and
 $\ket{j_i}$ is the $j_i$th standard basis vector of $\CC^k$.
 }
We also need the following lemmas.
\lemu{[See \cite{wocjan2004new}] \label{lemm_mub}
Let $\{m_{11},\dots,m_{1x},m_{21},\dots,m_{2x},\dots,m_{n1},\dots,m_{nx}\}$ be a $(n,x)-$net
 and $H$ an arbitrary generalized Hadamard matrix of size $x$
 (all its entries have modulus one and $HH^*=xI_x$ ).
Then the $n$ sets for $b=1,\dots,n$
$$L_b:=\{\frac{1}{\sqrt{x}}(h_l\uparrow m_{bi})\mid l=1,\dots,x, i=1,\dots,x\}$$
are $n$ MUBs for the Hilbert space $\CC^k$.
}
\lemu{[See \cite{wocjan2004new}] \label{lemm_mol}
The existence of $w$ MOLS is equivalent to the existence of a $(n,x)$-net with $n=w+2$.
}
By lemma \ref{lemm_mub} and Lemma \ref{lemm_mol} we know that
 there exist $N_{\text{MOLS}}(x)+2$ MUBs for the Hilbert space $\CC^k$.
On the other hand, there are many results on the value of $N_{\text{MOLS}}(x)$.
 A table of $N_{\text{MOLS}}(x)$ for $x<10000$ is presented in \cite{abel1996mutually},
 and for $x$ large enough, there is a bound $N_{\text{MOLS}}(\sqrt{k})+2\ge k^{1/29.6}$
 by \cite{wocjan2004new}.

In the following theorem, we show how to use mutually orthogonal Latin square to construct more MUBs in bipartite system.

\thm{\label{thm_sq}
Let $d$ be an odd number.
Write $d=p_1^{a_1}\dots p_s^{a_s}$ with $p_1^{a_1}\le \dots\le p_s^{a_s}$.
Suppose  $k=x^2$  is a square number. Then
\eq{\label{eq_mols}
M(d,kd)\ge \min\{N_{\text{MOLS}}(\sqrt{k})+2,M(d,d)\},
}
 where $N_{\text{MOLS}}(\sqrt{k})+2\ge k^{1/29.6}$ for all $\sqrt{k}$ but finitely many exceptions.
}
\pf{
Using the same discussion as in the proof of Theorem \ref{thm_odd} and
replacing $B_0,\dots, B_{q'_1}$ by $L_1,L_2,\dots,L_{N_{\text{MOLS}}(x)+2}$,
 we obtain the bound \eqref{eq_mols}.
For the lower bound on $N_{\text{MOLS}}$, we refer to \cite{wocjan2004new}.
}

Now, we compare the bounds obtained by Latin square method with reducing prime power method, we obtain many interesting results. In some cases, we find by Latin square, we can reach greater bounds than reducing into prime power problem.

\rk{
In the following  cases, Theorem \ref{thm_odd} is better than Theorem \ref{thm_sq}:

\enmt{[(1)]
\item Obviously, $k$ is not square, but is an odd number or a prime power.
\item $k=p^{2e}$, where $p$ is an arbitrary prime and $e\ge 1$.
 Theorem \ref{thm_sq} gives  $M(d,kd)\ge \min\{p^e+1,M(d,d)\}$,
  but Theorem \ref{thm_odd} gives $M(d,kd)\ge \min\{p^{2e}+1,M(d,d)\}$.
\item $x=76$, then $N(k=x^2)\ge 2^4+1=17$ and $N_{\text{MOLS}}(\sqrt{k}=76)\ge 6$ by \cite{wocjan2004new}.
 Theorem \ref{thm_sq} gives  $M(d,kd)\ge \min\{8,M(d,d)\}$,
  but Theorem \ref{thm_odd} gives $M(d,kd)\ge \min\{17,M(d,d)\}$.
}

However, in some cases, Theorem \ref{thm_sq}
 is better than Theorem \ref{thm_odd}:

\enmt{[(1)]
\item $k\ge 2$ is square number, i.e.,
$k=(p'_1)^{2e_1}\dots (p'_l)^{2e_l}=x^2$ with $(p'_1)^{2e_1}\le \dots\le (p'_l)^{2e_l}$.
\enmt{
\item $x\equiv 2 \pmod 4$. Then the minimal prime power dividing $x$ is 2.
 Thus Theorem \ref{thm_odd} gives $M(d,kd)\ge \min\{5,M(d,d)\}$.
By Beth's result \cite{beth1999design}
we know that $N_{\text{MOLS}}(x)\ge 6$ for $x\ge 76$.
Therefore, Theorem \ref{thm_sq} gives $M(d,kd)\ge \min\{8,M(d,d)\}$ for $k=x^2$, $x\ge 76$.
\item $k=26^2=2^2\times 13^2$. We have $(p'_1)^{a'_1}+1=5$ and $N_{\text{MOLS}}(\sqrt{k}=26)+2\ge 6$
where $N_{\text{MOLS}}(26)\ge 4$ (c.f. \cite{wocjan2004new}).
It follows from Theorem \ref{thm_odd} that \par
$M(d,kd)\ge \min\{5, M(d,d)\}$
, and by Theorem \ref{thm_sq}, then $M(d,kd)\ge \min\{6,M(d,d)\}$. Hence Theorem \ref{thm_sq} is better.

\item $l\ge 35$. Then  we have $N_{\text{MOLS}}(\sqrt{k})+2\ge k^{\frac{1}{29.6}}\ge(p'_1)^{2e_l}+1=(p'_1)^{a'_1}+1$.
Actually, since for $k$ large enough,
$$N_{\text{MOLS}}(\sqrt{k})+2\ge k^{\frac{1}{29.6}}=
 [(p'_1)^{2e_1}\dots (p'_l)^{2e_l}]^{\frac{1}{29.6}}
 \ge [(p'_1)^{2e_1}]^{\frac{l}{29.6}},$$
it suffice to show that $[(p'_1)^{2e_1}]^{\frac{l}{29.6}}\ge (p'_1)^{2e_1}+1$.
This is the case when $l\ge  35\ge 29.6\log_{(p'_1)^{2e_1}}((p'_1)^{2e_1}+1)$,
according to $\log_{(p'_1)^{2e_1}}((p'_1)^{2e_1}+1)\le \log_4 5$.
}
\item $k$ is not square, but $k=26^2p_1^{a_1}\dots p_n^{a_n}$ where $p_i \ge 3$.
For instance, $k=26^2\times 5$.
}}

In the previous remark, we  exhibit some examples to compare
 the two lower bounds given by Theorems \ref{thm_odd} and \ref{thm_sq}.
In all cases, these two theorems together give the combined lower bound:
\coru{
Let $d$ be an odd number.
Write $d=p_1^{a_1}\dots p_s^{a_s}$ with $p_1^{a_1}\le \dots\le p_s^{a_s}$.
Suppose that $k=(p'_1)^{2e_1}\dots (p'_l)^{2e_l}$ is a square number with $(p'_1)^{2e_1}\le \dots\le (p'_l)^{2e_l}$.
Then
$$M(d,kd)\ge \min\{ \max\{N_{\text{MOLS}}(\sqrt{k})+2,(p'_1)^{a'_1}+1\},M(d,d)\}.$$
}

\rk{If $d=2^m$, our bounds of $M(d,kd)$ still hold.
The proof is same as the argument
 except that we shall use the $2(2^m-1)$ MUMEBs constructed in
 \cite{xu2017construction}.
}

\section{Conclusion}\label{sec_con}
In this paper, we study the constructions of MUMEBs
 in bipartite system $\ckd$ for general $k$ and odd $d$.
 First, by using properties of Gauss sums, we construct $2(p^a-1)$ MUMEBs in $\cd$ for arbitrary odd $d$. It improves the known lower bound $p^a-1$ for odd $d$ and it also generalizes the lower bound $2(p^a-1)$ for $d$ being a single prime power.
Then, we construct MUMEBs in $\ckd$ for general $k\geq 2$ and odd $d$. We get the similar lower bounds as $k,b$ are both single prime powers.
At last, when $k$ is a square number, by using mutually orthogonal Latin squares, we can construct more MUMEBs in $\ckd$, and obtain greater lower bounds than reducing the problem into prime power dimension in some cases. Certainly, the above bounds of $M(d,kd)$ still hold for $d=2^m$. In the future work, we will consider the construction problem of MUMEBs in bipartite system $\ckd$ for general $d$.

%
%
%
%
%

\section{Acknowledgment}
This work was partially supported by National Key Research and Development Program of China under grant 2016YFB1000902, National Research Foundation of China (Grant No.61472412), and Program for Creative Research Group of National Natural Science Foundation of China (Grant No. 61621003).

\section{References}


\end{document}